\documentclass[english]{article}
 
\usepackage[utopia]{mathdesign}
\usepackage[T1]{fontenc}
\usepackage[latin9]{inputenc}
\usepackage[a4paper]{geometry}
\usepackage{booktabs}
\usepackage{mathrsfs}
\usepackage{amsmath}
\usepackage{amsthm}
\usepackage{graphicx}
\usepackage[authoryear]{natbib}
\usepackage[small]{titlesec}
\usepackage[font=small,labelfont=bf]{caption}

\usepackage{babel}

\makeatletter
\@ifundefined{date}{}{\date{}}
\makeatother

\begin{document}

\title{Should transparency be (in-)transparent? On monitoring aversion and cooperation in teams\thanks{We gratefully acknowledge financial support from the University of Birmingham. J. Jarke-Neuert acknowledges financial support from the Department of Socioeconomics at University of Hamburg and the Deutsche Forschungsgemeinschaft (DFG) under project number EXC 2037 390683824.}}
\author{Michalis Drouvelis\thanks{Department of Economics, University of Birmingham, Edgbaston, B15 2TT, and CESifo, Munich, Germany.}\qquad{}Johannes Jarke-Neuert\thanks{Center of Earth System Research and Sustainability (CEN), University of Hamburg, Grindelberg 5, 20114 Hamburg, Germany. Correspondence to  johannes.jarke-neuert@uni-hamburg.de.}\qquad{}Johannes Lohse\thanks{Department of Economics, University of Birmingham, Edgbaston, B15 2TT.}}

\maketitle

\begin{abstract}
Many modern organisations employ methods which involve monitoring of employees' actions in order to encourage teamwork in the workplace. While monitoring promotes a transparent working environment, the effects of making monitoring itself transparent may be ambiguous and have received surprisingly little attention in the literature. Using a novel laboratory experiment, we create a working environment in which first movers can (or cannot) observe second mover's monitoring at the end of a round. Our framework consists of a standard repeated sequential Prisoner's Dilemma, where the second mover can observe the choices made by first movers either exogenously or endogenously. We show that mutual cooperation occurs significantly more frequently when monitoring is made transparent. Additionally, our results highlight the key role of conditional cooperators (who are more likely to monitor) in promoting teamwork. Overall, the observed cooperation enhancing effects are due to monitoring actions that carry information about first movers who use it to better screen the type of their co-player and thereby reduce the risk of being exploited.
\bigskip{}
\end{abstract}

\section{Introduction\label{sec:Introduction}}

\emph{Transparency} --- defined as the degree to which it is possible to observe who does what and when --- is believed to play a key role for cooperation within teams, coalitions, and organizations \citep[e.g.,][]{kramer1995trust, mcallister1995affect,dunphy1996teams, jones1998experience, Salas_Work, de2010does}.\footnote{The management literature uses transparency as an umbrella term for all forms of information acquisition and production that occur in an organization and aim at making information available to others without hierarchical structures determining the direction of the information flow. This differs from other commonly studied forms of transparency such as surveillance (i.e. supervision by managers in a hierarchical relationship) and disclosure (i.e. the active sharing of previously secret information by managers); for further reflections on this terminology see \citet{bernstein2017making}.} One natural mode of transparency production is \emph{monitoring}. Experimental and field evidence is consistent with the tenet that monitoring through an external agent supports cooperation or abstention  from opportunism, and in turn team output \citep[e.g.,][]{Nagin_Monitoring, dickinson2008does, Costa_SC, rustagi2010conditional, Aoyagi_Monitor}.\footnote{Note, however, that monitoring can also have adverse effects \citep{Orr_Shame, falk2006hidden}. We extend on this negative role of monitoring below.} Yet, monitoring within teams routinely comes in the \emph{endogenous} form of peer monitoring \citep{Salas_Work, Ferrin_Trust, Wolitzky_Network, goeschl2017trust, Li_IT}. That is, team members decide how much monitoring to apply to whom and when. Endogenous monitoring gives rise to another layer of transparency in which monitoring decisions itself may be transparent---a setting that we call \emph{transparent transparency}---or not. In this paper, we report the results of a novel behavioral lab experiment exploring whether two-person teams interacting repeatedly in a sequential Prisoner's Dilemma can obtain more efficient outcomes under transparent transparency versus the alternative mode of hidden monitoring.  

The sequential Prisoner's Dilemma is a generic model of situations in which team members have an incentive to behave opportunistically to the detriment of successfully completing a joint project, such that voluntary cooperation (i.e. abstaining from opportunism) of team members is essential for efficient outcomes \citep{jones1984task}. For instance, when team members supply different inputs, a project's overall success may depend on the (successive) effort supplied by each team member. However, often individual team members may benefit from shirking and providing less than optimal effort to save on private opportunity costs. If some or all team members can monitor the actions of other team members, then contribution decisions can be made contingent on earlier observed actions, which provides a way for implicit sanctions. Yet, little is known about whether endogenous opportunities for monitoring within a team can enhance or hinder team performance when monitoring itself is transparent or not. Our study aims at filling this gap in the literature. 

Economic models of workplace relationships describe monitoring as essential for regulating opportunistic behaviors such as shirking and reducing them via implicit or explicit sanctions. This is commonly referred to as the \emph{disciplining effect} of monitoring \citep{Nagin_Monitoring,dickinson2008does}. However, the act of monitoring could itself have further effects on the behavior of the person being observed that diminish or strengthen its disciplining effect. In theory, two different behavioral mechanisms appear equally plausible.

First, information gathered through monitoring is more valuable to team members who plan to instrumentally use it by making their own effort choice contingent on the earlier effort choices of the other team member. The use of monitoring hence reveals the intentions of the person making use of it. In other words, making monitoring observable reduces uncertainty by allowing to discriminate between team members who condition their effort on the actions of the other team members, and those who do choose their effort independently. This information reduces the risk of being exploited and hence increases the propensity of mutual cooperation. If this effect is anticipated, it should result in a higher propensity to monitor in the first place. We call this the ``screening mechanism''. 

Second, active monitoring might be perceived as a signal of distrust or an attempt to exercise control over the person being monitored. There is mounting evidence documenting that individuals dislike being tightly controlled---a phenomenon that has been termed \emph{control aversion} \citep[e.g.,][]{falk2006hidden,boly2011incentive, ziegelmeyer2012hidden,belot2016spillover}. Such aversion has been shown to trigger negative reciprocity towards the person exercising control. If this mechanism dominates, monitoring---when observed---would lead to an erosion of trust and cooperation in repeated interactions and secret monitoring would hence be a preferable organizational practice. Similarly, the anticipation of control aversion could in turn reduce the incentives to make use of the monitoring option in the first place \citep{frey1993does}. Following \citet{falk2006hidden}, we call this the ``cost of control mechanism''. 

Both mechanisms predict opposing effects on cooperation, and possibly opposite second-order effects on monitoring behavior. From the extant literature, it is yet unclear which of these two effects might dominate and most importantly, which mechanism could explain the potentially differential impact on cooperative behavior. 

We conduct a laboratory experiment specifically designed to discriminate between the two mechanisms. As a generic and parsimonious representation of an ongoing teamwork setting, we use the well-known repeated two-player sequential Prisoner's Dilemma \citep{clark2001sequential}. This game  captures situations where team performance is highest when both team members put in high effort, but where each individual team member has an incentive to put in low effort. We extend the sequential Prisoner's Dilemma by including a monitoring opportunity rather than revealing the effort choice of the first-moving team member automatically. Monitoring allows the second-moving team member to observe the prior effort choice of the other team member before deciding about her own effort. Without monitoring, the choice of the other team member is only revealed after both have made their decisions. The monitoring opportunity is asymmetric: the second-moving team member can monitor the first-moving team member, but not vice versa. The asymmetric structure of the game allows for a sharper analysis of behavior and motivational structure compared to a symmetric game.

In this basic setting, we compare behavior in a treatment condition where the act of monitoring itself is observable at the end of an interaction (condition TT for ``transparent transparency'') to a treatment condition in which the act of monitoring is not observable (condition IT for ``in-transparent transparency''). We also implement two further baseline conditions that are identical to TT and IT except that monitoring choices are made by the computer (instead of the second-moving team members) using the same probabilities with which real second movers monitored in TT and IT. In the control conditions, which we label TTex and ITex, any strategic function of monitoring is muted, such that first movers cannot reasonably infer any information regarding their partner from observed monitoring patterns.

Contrasting experimental conditions in which monitoring decisions are observable (TT and TTex) at the end of an interaction with those where they are not observable (IT and ITex) is informative with respect to the two mechanisms outlined above. Both the ``screening mechanism'' and the ``cost of control mechanism'' require monitoring to be observable. Predictions about monitoring choices and cooperative behavior are orthogonal under both mechanisms, such that observed behavior patterns reveal if either of the two mechanisms dominates. Specifically, contrasting conditions with exogenous (TTex and ITex) and endogenous (TT and IT) monitoring actions is informative with respect to the "cost of control mechanism", since the second mover cannot be made responsible for randomly determined monitoring choices, such that the ``cost of control mechanism'' is muted.

Our experimental results suggest that transparent transparency is indeed a superior condition for teamwork. When monitoring choices are observable, there is more mutually beneficial cooperation within teams. This is especially true for the early stages of an interaction, in which uncertainty about the other team member's type is largest. Our findings hence speak against significant ``costs of control'' but support the screening mechanism that allows team members to better understand the type of their partner. We believe that the repeated interaction is pivotal for this result, and could explain the difference to seminal experimental evidence of ``costs of control'' that is obtained from one-shot interactions \citep{falk2006hidden}.

The results inform recent discussions in the literature on optimal management practices. We offer direct evidence on how different transparency regimes affect team performance. This adds to a larger literature on the question of how to improve team performance by thinking carefully about individual and team incentives \citep{chen2013should,lount2014working,bradler2016employee} or through paying closer attention to team composition and diversity \citep{apesteguia2012impact,hoogendoorn2013impact}. By contrasting two theoretically plausible mechanisms, our results are also informative with regard to basic motives that underpin decisions to cooperate in settings where opportunistic behavior is in the material self-interest of decision makers. While the literature typically links cooperative behavior to preference-based explanations \citep[e.g.,][]{fehr2002social,fischbacher2010social}, our results show that decisions to cooperate in teams may also reflect the perceived intentions of cooperation partners  \cite[e.g.,][]{falk2008testing, von2013intention, rand2015s}. 

Throughout the remainder of the paper, we proceed as follows. In Section \ref{sec:Experiment}, we describe the experimental design and procedures. The results are exposed in Section \ref{sec:Results}. Section \ref{sec:Mechanisms} discusses possible mechanisms, and in Section \ref{sec:Conclusion} we conclude with some implications for organizational management practice. 

\section{Experiment\label{sec:Experiment}}
In this section, we describe the design and procedures of the experiment that is organized around the well-known repeated two-player sequential Prisoner's Dilemma \citep{clark2001sequential}, which we frame here as a ``teamwork game''. To the participants, the game was described in neutral terms. The full instructions are provided in the Appendix.\footnote{Experimental data and analysis files are available on the Open Science Framework (OSF) under DOI 10.17605/OSF.IO/D5Q7G.}

\subsection{Decision task\label{subsec:Task}}
Participants were randomly allocated to one of the two roles, player $A$ (first mover) or player $B$ (second mover) and remained in the same role throughout the experiment. The experiment consisted of a 15-round repetition of the following stage game. At the beginning of a round $t$ player $A$ chooses between action ``low effort'' and action ``high effort''. Player $B$ either observes player $A$'s effort choice or not---depending on the monitoring structure described below---and then also chooses between action ``low effort'' and action ``high effort''. The payoffs are shown in table \ref{tab:Payoffs}. They are substantively identical to the ``baseline game'' in \citet{clark2001sequential}.
\begin{table}[h!]
\centering{}\caption{Stage game payoffs.\label{tab:Payoffs}}
\begin{tabular}{lcc}
\toprule 
 & high effort  & low effort \tabularnewline
\midrule
high effort  & $\left(8,8\right)$ & $\left(0,10\right)$\tabularnewline
low effort  & $\left(10,0\right)$ & $\left(2,2\right)$\tabularnewline
\bottomrule
\multicolumn{3}{p{.5\textwidth}}{\scriptsize{\textit{Table notes:} Payoffs are in pounds sterling. The left hand number is the payoff of the row player, the right hand number is the payoff of the column player.}}
\end{tabular}

\end{table}

Given these payoffs, both players have an incentive to exert low effort.\footnote{Yet, if the first mover believes for some reason that the second mover will put in high effort only in response to high effort (i.e. that the second mover is a conditional cooperator), it is also in the former's material interest to provide high effort. But conditional cooperation is apparently inconsistent with the second mover's material interests.} Mutual cooperation (i.e. both players exerting high effort) maximizes group payoffs, such that there is a tension between individual and collective interest.

The experiment was framed neutrally: the high effort choice was called action ``yellow'' and the low effort choice was termed action ``pink''. We observed play of the game in four experimental conditions.

\subsection{Conditions\label{subsec:Conditions}}
In the endogenous monitoring conditions (TT and IT), second movers decided if they wish to monitor and learn the first mover's decision prior to submitting their own action. Thereafter, they choose between ``high effort'' or ``low effort'' for themselves. In the TT condition, the monitoring choice is revealed to the first mover on a feedback screen at the end of a round. In the IT condition, this information is never revealed. The contrast between those two conditions allows us to test the main question of the paper i.e., whether making transparency itself transparent in repeated interactions is conductive to team performance.

The two exogenous monitoring conditions (TTex and ITex) differ from these procedures only by removing the control over the monitoring action from second movers. Instead, monitoring was exogenously determined. To make the comparison to the TT and IT conditions as meaningful as possible, we first completed the TT and IT sessions and then set the probability (resolved by the computer) to be monitored faced by an average first mover in each round equal to the relative frequencies of monitoring observed in TT and IT, respectively. As in TT and IT, the exact probability was unknown to both participants in the group ex ante. Participants were, however, made aware that the probability would reflect average choices of participants who had taken part in a previous experimental session. This information allows first movers to form expectations regarding the probabilities to be monitored, while leaving the exact probabilities unknown, as in the endogenous monitoring treatments. 

After both players in a group submitted their choices, there was a feedback screen (at the end of each round) showing to both group members the corresponding outcomes. This means that the team members were fully informed about the history of effort choices at the end of each round. Hence, the TT and IT (TTex and ITex) treatments only differ in terms of the information that the first mover received regarding the monitoring choice, but not in terms of what both parties can observe regarding the action taken in each round. The intention behind this design feature was to reflect how real teamwork usually works: effort choices (inputs) are generally not readily observable, but outcomes are revealed at the end of a project. This leaves the full scope of finitely repeated game strategies (e.g., reputation building, trigger strategies) intact while only blocking or opening the channel of information transmission through the monitoring choice \citep{andreoni1993rational}. 

After the main part of the experiment, we elicit one further choice using a strategy method design, i.e. we ask first movers (player $A$) to submit an unconditional choice and second movers (player $B$) to submit a decision plan that is conditional on A's choice. We will use these plans to classify second movers according to their preferred plan into conditional and unconditional cooperators (defectors). For details, see section \ref{subsec:TaskSup} of the Appendix.

\subsection{Procedures\label{sec:Procedures}}
The experiment was conducted at the Birmingham Experimental Economic Lab (BEEL) in spring 2018. 

There were ten sessions with 16-20 subjects per session for a total of 176 participants recruited from the standard pool of student subjects using ORSEE \citep{Greiner_ORSEE}. Randomization was done at the session level. We had 56 subjects in IT, 52 in TT, 34 in ITex, and 34 in TTex. The subjects were from mixed disciplines, including economics (24\%). There was a nearly balanced ratio of female (56\%) to male (44\%) participants. Subjects took part in no more than one session of the experiment (between subjects design). 

Upon arrival, participants were seated at a randomly allocated computer terminal. The experimenter made a short set of general announcements before participants began with the computerized part of the experiment, which was programmed in z-Tree \citep{Fischbacher_zTree} and included
all remaining instructions (available in the supplementary materials). After reading the instructions, participants were given the opportunity to ask clarifying questions in private. All other forms of communication were forbidden. 

After completing the task, participants filled out a short set of questions on demographic attributes and personality measures. At the end of the experiment, one round was selected randomly for anonymous payment that was made in cash. All sessions lasted for approximately 60 minutes and participants earned GBP 8.90 (Min: GBP 2.50, Max: GBP 13.50) including a show-up fee of GBP 2.50.

\section{Main results\label{sec:Results}}
We first analyze the monitoring choices across treatments and then examine how team performance is affected by the different treatment manipulations.

\subsection{Monitoring \label{subsec:MonitoringAll}}
\begin{figure}
\centering{}\caption{Relative frequency of monitoring (TT vs. IT)\label{fig:ExpMonitoring}}
\includegraphics[width=0.75\textwidth]{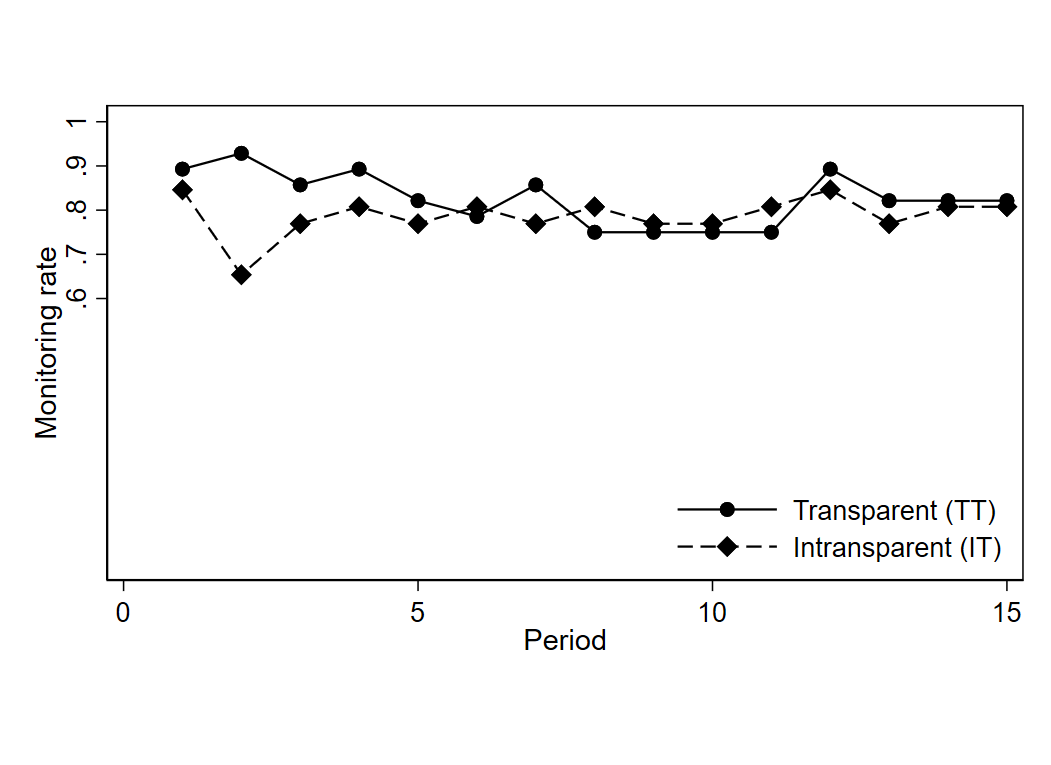}
\end{figure}

Averaging across all periods, we do not observe significant differences in monitoring patterns. Monitoring is high but not perfect in both the TT (79\%) and the IT (83\%) condition.\footnote{By design, in the exogenous monitoring treatments, the monitoring rates are chosen to be similar with 80\% for the ITex treatment and 83\% for the TTex treatment.} Figure \ref{fig:ExpMonitoring} displays the evolution of the monitoring rate over the 15 periods in both endogenous monitoring conditions. Apart from period 2, there are no clear differences between treatments and the monitoring rate stabilizes at approximately 80\% throughout the progression of the repeated game.

Table \ref{tab:ExpMonitoring} provides a breakdown of the relative frequency of monitoring by observability of the monitoring choice for different period intervals. Monitoring is slightly lower in early periods of the TT treatment, but there is strong convergence in later periods. For periods 1-5 the average rate is 12 percentage points lower, a difference that is, however, not significant at conventional levels (Mann-Whitney test, $p=.31$). This difference mainly reflects the strong drop of monitoring in the second period. 

All remaining comparisons in Table \ref{tab:ExpMonitoring} confirm that there are no significant differences in monitoring rates across treatments. The absence of differences in the monitoring frequencies implies that any potential differences in cooperation are not due to differences in monitoring but can be fully attributed to our exogenous variation of monitoring observability.
\begin{table}
\centering{}\caption{Relative frequency of monitoring by observability of the monitoring
choice (TT vs IT)\label{tab:ExpMonitoring}}
\begin{tabular}{lccccc}
\toprule 
 & \multicolumn{5}{c}{Period}\tabularnewline
\cmidrule{2-6} 
Treatment & 1 & 1-5 & 6-10 & 11-15 & 15\tabularnewline
\midrule 
IT & 0.89 & 0.88 & 0.77 & 0.82 & 0.82\tabularnewline
TT & 0.83 & 0.76 & 0.78 & 0.80 & 0.80\tabularnewline
\bottomrule
\end{tabular}
\end{table}

Overall, our findings regarding the monitoring choices of second movers provide evidence that the average second mover does not condition her monitoring behavior on whether that behavior is observable by the first mover or not.

\subsection{Team performance}

We next compare average group-level effort provision when monitoring choices are observable (TT and TTex) against when they are hidden (IT and ITex). All statistical tests in this section are Mann-Whitney tests applied to group-level averages over time. 

We find the relative frequency of high effort choices to be 15 percentage points higher ($p=.056$) in the TT and TTex conditions (72\% pooled) relative to the IT and ITex conditions (58\% pooled). Moving from an endogenous (TT and IT) to an exogenous (TTex and ITex) monitoring environment, we observe that the frequency of high effort choices is 68\% (pooled) and 60\% (pooled), respectively. Yet, this 8 percentage point drop in high effort choices is not statistically significant ($p=.273$). 

\begin{figure}
\centering{}\caption{Relative frequency of high effort provision (high effort rate) over time by treatment condition\label{fig:ExpCooperation}}
\includegraphics[width=0.75\textwidth]{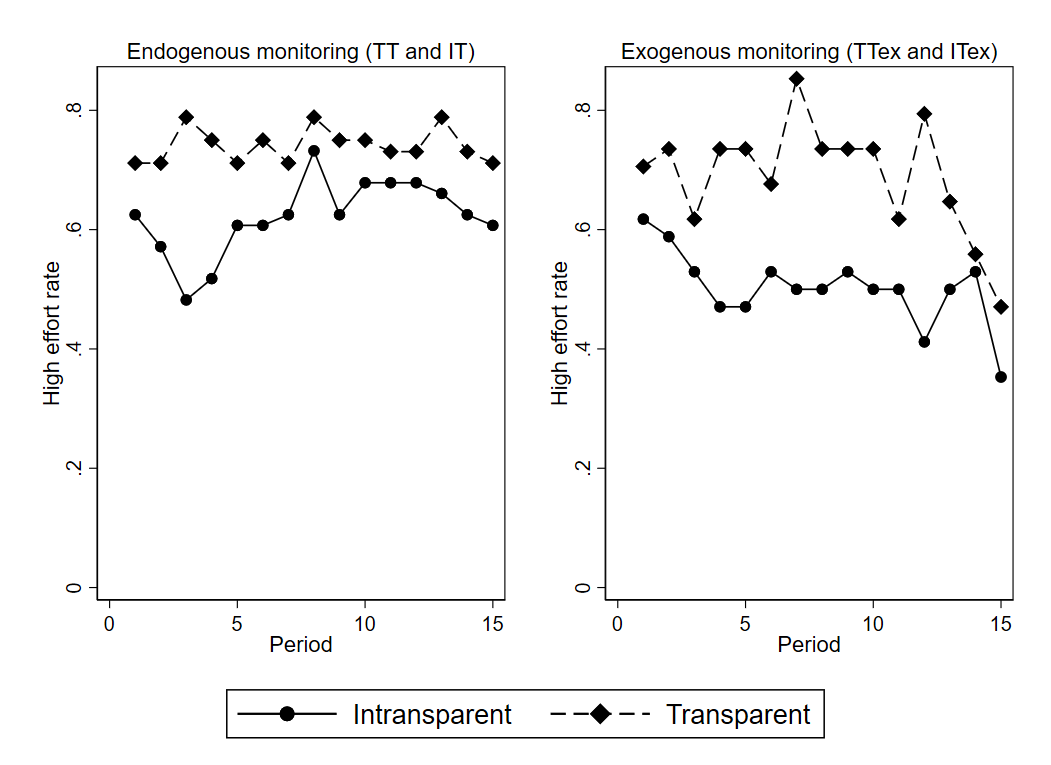}
\end{figure}

Figure \ref{fig:ExpCooperation} displays the dynamics of team performance across our four treatment conditions. Comparing choices across all 15 periods for the endogenous monitoring treatments, we continue to find lower levels of high effect choices in the IT condition (62\%) compared to the TT condition (74\%), which is not significant ($p=.125$). Similarly, we find lower levels of high effort choices in the ITex condition (50\%) compared to the TTex condition (69\%) which are not significantly different either ($p=.1928$). 

Table \ref{tab:ExpCooperation} summarizes the frequency of high effort choices in each of the treatments. In the endogenous monitoring conditions, the rate of cooperation is lower in each period for the IT treatment compared to the TT treatment. The difference is larger in the first five periods (TT vs. IT: 73 \% vs 56 \%; $p=.0614$) and becomes less pronounced and statistically insignificant in the later periods. There is only a small end-game effect. 

Similarly, in the exogenous monitoring treatments, we find lower rates of high effort choices in the ITex treatment (50\%) compared to the TTex treatment (69\%). The dynamics of effort provision follow a different pattern, though. Here, we find similar rates of cooperation in the first five periods ($p=.21$), yet more pronounced differences in periods 6-10 ($p=.0950$). 

There is also a stronger end game effect in both of the exogenous monitoring treatments. Comparing the TT and the TTex treatments, we find that in the early rounds of the game there is little difference (73\% and 71\%, respectively). Only in the last rounds there is a more noticeable difference which is due to the significant difference in the strength of the end-game effect ($p=.0930$). The high effort rates also evolve in similar patterns in the early rounds of the IT and ITex treatments with larger differences in later periods including a more significant end-game effect in the ITex treatments ($p=.0861$). 

\begin{table}
\caption{Maximum likelihood estimates and standard errors of estimates (SEEs)
of the average partial effect (APE) of observable vs. non-observable
monitoring on group-level effort choice patterns\label{tab:Reg1}}
\begin{centering}
\begin{tabular}{@{\extracolsep{\fill}}lrrrrr}
\toprule 
 & \multicolumn{2}{c}{Mixed-effects logit} &  & \multicolumn{2}{c}{Mixed-effects probit}\tabularnewline
\cmidrule{2-3} \cmidrule{3-3} \cmidrule{5-6} \cmidrule{6-6} 
 & Estimate & SEE ($p$) &  & Estimate & SEE ($p$)\tabularnewline
\midrule
Two low & $-.1252$ & $.0519$ $\left(.016\right)$ &  & $-.1309$ & $.0563$ $\left(.020\right)$\tabularnewline
One low, one high & $-.0350$ & $.0150$ $\left(.020\right)$ &  & $-.0321$ & $.0126$ $\left(.011\right)$\tabularnewline
Two high & $.1601$ & $.0653$ $\left(.014\right)$ &  & $.1629$ & $.0684$ $\left(.017\right)$\tabularnewline
\bottomrule
\end{tabular}
\par\end{centering}
\textit{\scriptsize{}\smallskip{}
}\\
\textit{\scriptsize{}Table notes:}{\scriptsize{} Shown are maximum
likelihood estimates of the average partical effect on the probability
of observing the respective group-level effort choice patterns in
an average round (i. e. the respective probability in the observable
monitoring condition minus the respective probability in the non-observable
monitoring condition, ceteris paribus) and SEEs derived via the delta
method from mixed-effects ordered logit and ordered probit regressions
with group-level random intercepts of a two-by-two treatment indicator
on the count of high effort choices. The $p$-value of a Wald test
of the null hypothesis that the respective estimate is equal to zero
is reported in parentheses. Rejections are ``highly significant''
for $p<.01$, ``significant'' for $p<.05$, and ``marginally significant''
for $p<.1$. Both regressions have $1,320$ observations clustered
in $88$ groups (i. e. $15$ rounds per group) and use mean-variance
adaptive Gauss--Hermite quadrature integration. Limit log-likelihood
was $-838.35149$ in the logit and $-837.32576$ in the probit model.
Likelihood-ratio tests reject the respective pooled specification
without random effects ($\bar{\chi}^{2}\left(01\right)=765.98$ and
$p=.0000$ for logit, $\bar{\chi}^{2}\left(01\right)=768.52$ and
$p=.0000$ for probit). The joint Wald tests reject the null model
($\chi^{2}\left(3\right)=8.09$ and $p=.0441$ for logit, marginally
for probit at $\chi^{2}\left(3\right)=7.58$ and $p=.0555$). Additional
session-level random intercepts are close to zero and not significant.}{\scriptsize\par}
\end{table}

These non-parametric results are complemented by Wald tests on mixed-effects ordered logit and probit regression estimates shown in Table \ref{tab:Reg1}. The dependent variable is the count of high effort choices (taking values 0, 1, and 2) in a given group and round. The models include group-level random intercepts. This method exploits more variance in the data than the non-parametric rank-order tests, while  maintaining the assumptions that observations across periods in a given group are allowed to be dependent while group-level outcomes are independent. The results provide further evidence that the observability of monitoring has a strong effect: while the observability treatment \emph{decreases} the probability of both players putting in low effort significantly by about 13 percentage points, and of only one player providing high effort by about 3 percentage points, it \emph{increases} the probability of both players putting in high effort significantly by more than 16 percentage points relative to the non-observability control condition, all else equal. The effect of moving from an endogenous to an exogenous monitoring setting is small in comparison and does not reach statistical significance for any of the outcome categories (the results are shown in Table \ref{tab:Reg2} in the Appendix).  

\begin{table}
\centering{}\caption{Relative frequency of cooperation by treatment condition\label{tab:ExpCooperation}}
\begin{tabular}{lccccc}
\toprule 
 & \multicolumn{5}{c}{Period}\tabularnewline
\cmidrule{2-6} 
Treatment & 1 & 1-5 & 6-10 & 11-15 & 15\tabularnewline
\midrule 
IT & 0.62 & 0.56 & 0.65 & 0.64 & 0.60\tabularnewline
TT & 0.71 & 0.73 & 0.75 & 0.73 & 0.71\tabularnewline
ITex & 0.62 & 0.54 & 0.51 & 0.45 & 0.35\tabularnewline
TTex & 0.71 & 0.71 & 0.75 & 0.61 & 0.47\tabularnewline
\bottomrule
\end{tabular}
\end{table}

In sum, our experimental results provide three key insights into how making transparency transparent affects team performance. First, independent of transparency, team members frequently make use of the possibility to monitor other team members. Second, transparent transparency is conducive to team performance by reducing the rates at which team members exploit possibilities for acting opportunistically. Third, as team members interact repeatedly, this effect is most pronounced in early interactions where team members have had little chance to observe the actions of other team members. Taken together, these three observations suggest that organizations would benefit from making transparency transparent. This is particularly true when the composition of teams changes frequently, such that team members would benefit more from making transparency transparent.

\section{Underlying mechanisms\label{sec:Mechanisms}}

In this section, we interpret our results in light of the two mechanisms we have outlined in the introduction. The analysis is organized around the idea that second movers differ in their cooperative type, that is, whether and for how many rounds they are willing to cooperate conditionally, which is consistent with previous evidence \citep{Cox_Reputations, Embry_FRPD}. Analogous to the formation of a new team, anonymous random matching at the beginning of the experiment implies that first movers are uncertain about the type of their partner. Observations of both effort and monitoring choices carry information for the first mover who can learn about the type of the second mover. A sophisticated second mover could even actively try to communicate her type through the combination of monitoring and effort choices when the former are observable. The key contrast between the TT and IT conditions is that one of the information transmission channels, namely monitoring actions, is selectively open (TT) or closed (IT). Hence, the information-carrier function of monitoring choices is selectively turned on in TT and off in IT.

\subsection{Monitoring rate}
The two mechanisms imply opposite incentives for second movers to monitor under TT vs IT. Assuming that second movers are aware of the possibility that first movers could use their monitoring choice to screen their cooperative type, the screening mechanism implies that the monitoring rate should be higher in the TT than in the IT condition. The cost of control mechanism on the other hand implies that second movers should be more reluctant to monitor in TT than in IT, because there is a hazard of an adverse response by the partner. In section \ref{subsec:MonitoringAll} we show that the monitoring rates do not differ between the two conditions, suggesting that any observed differences in cooperation rates cannot be attributed to monitoring rates.

In a complimentary approach, we connect second movers' monitoring behavior to their choices in a supplemental part of the experiment that we describe in the appendix. In this part, we ran a one-shot version of the stage game after the main decision task in all conditions. In this one-shot version, we elicited choices via the strategy method. Based on their preferred decision plan in the supplementary part, we can classify second movers into conditional cooperators, unconditional cooperators, and unconditional defectors. The screening mechanism suggests that the monitoring option should be chosen most frequently by subjects classified as conditional cooperators, because they have two separate reasons to do so: to condition their effort on the effort of the first mover and to signal their type. In line with these considerations, we find suggestive evidence that conditional cooperators make use of the monitoring option more often (88\%) than those  subjects preferring unconditional strategies in the supplemental part (71\%) (Mann-Whitney test, $p = .0902$). This evidence---although supporting the conclusions from the main part of the experiment that we discuss below---is only exploratory and needs to be considered carefully, because the supplemental part was conducted after the main experiment and may therefore be not entirely independent of second mover's experiences in the main part of the experiments.

\subsection{Effort provision}

\begin{figure}
\centering{}\caption{Relative frequency of high effort provision (high effort rate) by second movers over time by treatment condition\label{fig:ExpCooperationSM}}
\includegraphics[width=0.75\textwidth]{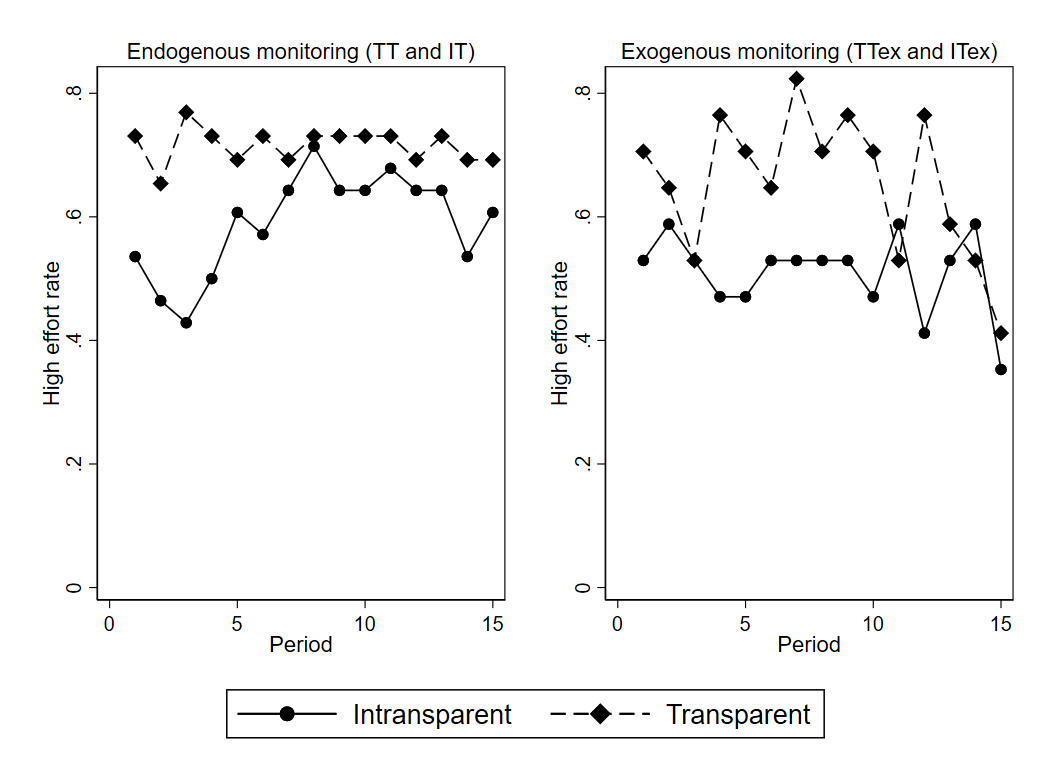}
\end{figure}

We first examine second movers' behavior over time in all treatments, as shown in Figure \ref{fig:ExpCooperationSM}. Comparing behavior across conditions in the left panel, we find higher levels of effort provision especially in TT compared to IT in the early rounds (Mann-Whitney tests: round 1: 73\% vs 53\%, $p=.139$; Round 1-5: 71\% vs. 50\%, $p=.0409$). There are no significant differences in effort provision in the second half of the game. Similarly, with exogenous monitoring, we find a higher frequency of effort provision in TTex than in ITex, which are most pronounced and significant in some of the middle periods of the game.\footnote{We note in passing that the end-game effects are stronger in the exogenous compared to the endogenous treatments (Mann-Whitney tests: IT vs. ITex, $p = .670$ and TT vs TTex $p = .0267$). While noteworthy, this difference does not speak to the mechanisms we are considering here.} The observation that second movers react to the observability of their monitoring decision (or realization in TTex) by cooperating more often suggests that they anticipate the screening effect.

Further support for this interpretation comes from an analysis of second mover behavior conditional on the (observed) choices of first movers.

Across all treatment conditions, averaging over time, we observe behavior strongly consistent with reciprocity (i.e. second movers respond with high effort to high effort, and with low effort to low effort). When observing first movers providing high effort, second movers reciprocate with high effort in 60 percent to 100 percent of the cases. Similarly, when observing low effort, second movers respond in-kind with low effort in 80 percent to 100 percent of the cases.

Table \ref{tab:ExpCooperationSMcond} summarizes reciprocal behavior across different period intervals. Making monitoring choices transparent has little effect on this reciprocal behavior with one exception: the top left panel indicates that subjects in the early rounds of the endogenous monitoring treatment are less inclined to reciprocate high effort when their monitoring choice is not observable to the first mover. This difference is weakly significant for the early periods (Pearson $\chi^2$ tests: round 1 $p=.082$; round 2 $p=.183$; round 3 $p=.078$) but disappears thereafter.\footnote{explanation from comment} This observation suggests that in early rounds, second movers in the IT condition realize that first movers cannot tell if their choices to provide low effort are conditional or unconditional. We provide further evidence for this interpretation when we examine first movers' behavior separately. 

\begin{table}
\centering{}\caption{Relative frequency of second mover's high effort choices conditional on the first mover's action by treatment condition\label{tab:ExpCooperationSMcond}}
\begin{tabular}{lcccccc}
\toprule 
 & \multicolumn{3}{c}{FM high effort} & \multicolumn{3}{c}{FM low effort}\tabularnewline
 & \multicolumn{3}{c}{Period} & \multicolumn{3}{c}{Period}\tabularnewline
\cmidrule{2-7} 
Treatment & 1-5 & 6-10 & 11-15 & 1-5 & 6-10 & 11-15\tabularnewline
\midrule 
IT & 0.68 & 0.88 & 0.86 & 0.13 & 0.14 & 0.10\tabularnewline
TT & 0.88 & 0.93 & 0.94 & 0.17 & 0.04 & 0.00\tabularnewline
ITex & 0.81 & 0.90 & 0.97 & 0.13 & 0.09 & 0.10\tabularnewline
TTex & 0.86 & 0.91 & 0.89 & 0.17 & 0.06 & 0.00\tabularnewline
\bottomrule
\end{tabular}
\end{table}

In a second step, we analyze how first movers are reacting to high effort and monitoring choices of second movers in the preceding round. Table \ref{tab:ExpCooperationFM} reports the average frequency of first-mover's high effort choices, split by different period intervals. For the endogenous monitoring condition, there is a slight indication that first movers are more cooperative in the TT as compared to the IT condition (Mann-Whitney test, $p=.1126$). Again, this difference is stronger in the first five periods (Mann-Whitney test, $p=.0983$) than in the remaining periods. In the exogenous monitoring treatments, despite starting at comparable cooperation rates in the first period, there is a difference between the TTex and ITex that persists until the final periods (group-average Mann-Whitney test, $p=.0666$). 

The main observation from this analysis indicates that first movers are more cooperative when they can observe second movers' monitoring choice in the previous period. This adds evidence that first movers' behavior is not driven by the control aversion mechanism but the screening mechanism. This interpretation is further supported by looking at first movers' behavior conditional on second movers' effort choice in Table \ref{tab:CoopFMCond}. 

\begin{table}
\centering{}\caption{Relative frequency of cooperation by first movers by treatment condition\label{tab:ExpCooperationFM}}
\begin{tabular}{lcccccc}
\toprule 
 & \multicolumn{5}{c}{Period} & \tabularnewline
\cmidrule{2-7} 
Treatment & 1 & 1-5 & 6-10 & 11-15 & 15 & Total\tabularnewline
\midrule 
IT & 0.71 & 0.61 & 0.66 & 0.67 & 0.60 & 0.65\tabularnewline
TT & 0.69 & 0.75 & 0.77 & 0.76 & 0.73 & 0.76\tabularnewline
ITex & 0.71 & 0.55 & 0.51 & 0.42 & 0.35 & 0.49\tabularnewline
TTex & 0.71 & 0.74 & 0.76 & 0.67 & 0.53 & 0.73\tabularnewline
\bottomrule
\end{tabular}
\end{table}

Independent of the treatment condition, first movers are more likely to provide high effort in the following round upon observing high effort by the second mover in the previous round. Low effort is also frequently reciprocated in-kind. However, first movers display a larger propensity to proposing a ``re-start'' by answering a previous low effort input with high effort, especially in the early rounds where this appears in more than 40\% of all occasions. There is weak evidence that first movers in the exogenous monitoring conditions continue with this behavior when monitoring choices are observable (Mann-Whitney test, period 6-10 $p=.0086$, period 11-15 $p=.0885$).
\begin{table}
\centering{}\caption{Relative frequency of high effort choices by first movers conditional on the second mover's previous action by treatment condition\label{tab:CoopFMCond}}
\begin{tabular}{lcccccc}
\toprule 
 & \multicolumn{3}{c}{SM high effort} & \multicolumn{3}{c}{SM low effort}\tabularnewline
 & \multicolumn{3}{c}{Period} & \multicolumn{3}{c}{Period}\tabularnewline
\cmidrule{2-7} 
Treatment & 1-5 & 6-10 & 11-15 & 1-5 & 6-10 & 11-15\tabularnewline
\midrule 
IT & 0.87 & 0.93 & 0.95 & 0.32 & 0.19 & 0.21\tabularnewline
TT & 0.89 & 0.98 & 0.98 & 0.44 & 0.24 & 0.21\tabularnewline
ITex & 0.77 & 0.84 & 0.70 & 0.22 & 0.15 & 0.12\tabularnewline
TTex & 0.93 & 0.90 & 0.91 & 0.39 & 0.39 & 0.28\tabularnewline
\bottomrule
\end{tabular}
\end{table}

We provide further evidence showing the positive effect of monitoring observability on effort provision patterns by looking at how first movers react to monitoring decisions.  In the endogenous monitoring treatments, we observe for most periods that those subjects who have been monitored previously are more likely to provide high effort in the next round, with an average difference of approximately 18 percentage points. We do not observe similar differences in the treatments where monitoring is applied exogenously. This implies that the monitoring choice itself has a screening value in that it makes first movers become more cooperative when they have observed that the second mover has decided to monitor their choice in the previous period.

In sum, these patterns are more suggestive of the screening mechanism than of the costs of control mechanism. Specifically, our data indicate that second movers are not more reluctant to use monitoring when their monitoring choice is observable. Moreover, their effort provision patterns suggest that they are aware of the signalling value of cooperating when looking. Similarly, first movers use the information carried by monitoring actions to screen between conditional cooperators and displaying an unconditional behavioural type (cooperation or defect) to mitigate the risk of being exploited. This in turn elevates their willingness to cooperate. By contrast, we find no evidence in favour of the control aversion mechanism, which would suggest lower rates of effort provision after observing the monitoring choices of others. 

\section{Conclusion\label{sec:Conclusion}}
We report an experiment testing for the behavioral effects of the observability of monitoring on the decision to monitor and the subsequent effect on cooperation. Our design allows us to disentangle two opposing effects that might be at work when there are possibilities to observe monitoring. On the one hand, monitoring actions could carry information about the acting player's type. On the other hand, the choice to monitor might be perceived as a signal of distrust or an attempt to exercise control. This can be attributed to hidden costs of control, which have been identified in different settings \citep{falk2006hidden,boly2011incentive}.

Our findings indicate that the observability of monitoring does not affect the frequency of monitoring, but significantly increases the frequency of cooperation. This offers evidence in favor of the screening mechanism, as the hidden benefits of the monitoring dominate the hidden costs of control.

For organizational practice, our results suggest that making transparency transparent is conducive for teamwork. Especially when teams are newly formed, but then interact repeatedly, observable monitoring reduces uncertainty, helps to identify cooperative interaction partners and thereby strengthens mutually beneficial cooperation. These considerations could inform a larger literature interested in improving team performance.

Our results also qualify the importance of ``costs of control''. While these have been found to be especially important in one-shot settings where strong control is exercised over the action space of agents \citep{falk2006hidden}, our findings suggest that ``costs of control'' are smaller when it comes to monitoring. There are two potential explanations that could be tested in future studies. It may simply be the case that monitoring is perceived as a less encroaching form of control than limiting the action space of agents. A second explanation could be that ``costs of control'' have been previously studied in one-shot settings. In our repeated setting, the long-term screening value of monitoring may dominate the short term cost of control.

\newpage
\bibliographystyle{apalike}
\bibliography{Jos_Bib2}

\newpage
\appendix

\section{Supplemental decision task\label{subsec:TaskSup}}
The following decision plans were available:  
\begin{enumerate}
\item Don't find out about Participant $A$'s choice and choose yellow 
\item Don't find out about Participant $A$'s choice and choose pink 
\item Find out about Participant $A$'s choice and choose yellow independently of $A$'s choice 
\item Find out about Participant $A$'s choice and choose pink independently of $A$'s choice 
\item Find out about Participant $A$'s choice and choose pink if $A$ chooses pink, and yellow if $A$ chooses yellow 
\item Find out about Participant $A$'s choice and choose yellow if $A$ chooses pink, and pink if $A$ chooses yellow 
\end{enumerate}
\newpage
\section{Additional Results \label{subsec:Additional Results}}

In this part of the appendix, we provide further information on regression based tests of our main results. Both the average partial effects discussed in the main body of the paper in table \ref{tab:Reg1} and the effects shown here (table \ref{tab:Reg2}) come from a mixed effects orderer logit (ordered probit) model that takes the group level outcome (both high effort, one high effort, none high effort) as the dependent variable. The independent variable is a treatment indicator for assignment to the observable monitoring (exogenous monitoring) arms of the experiment.

\begin{table}[h!]
\caption{Maximum likelihood estimates and standard errors of estimates (SEEs)
of the average partial effect (APE) of endogenous vs. exogenous monitoring
on group-level effort choice patterns\label{tab:Reg2}}
\begin{centering}
\begin{tabular}{@{\extracolsep{\fill}}lrrrrr}
\toprule 
 & \multicolumn{2}{c}{Mixed-effects logit} &  & \multicolumn{2}{c}{Mixed-effects probit}\tabularnewline
\cmidrule{2-3} \cmidrule{3-3} \cmidrule{5-6} \cmidrule{6-6} 
 & Estimate & SEE ($p$) &  & Estimate & SEE ($p$)\tabularnewline
\midrule
Two low & $.0727$ & $.0550$ $\left(.187\right)$ &  & $.0761$ & $.0604$ $\left(.207\right)$\tabularnewline
One low, one high & $.0179$ & $.0152$ $\left(.237\right)$ &  & $.0171$ & $.0125$ $\left(.173\right)$\tabularnewline
Two high & $-.0906$ & $.0694$ $\left(.192\right)$ &  & $-.0932$ & $.0723$ $\left(.197\right)$\tabularnewline
\bottomrule
\end{tabular}
\par\end{centering}
\textit{\scriptsize{}\smallskip{}
}\\
\textit{\scriptsize{}Table notes:}{\scriptsize{} Shown are maximum
likelihood estimates of the average partical effect on the probability
of observing the respective group-level effort choice patterns in
an average round (i. e. the respective probability in the observable
monitoring condition minus the respective probability in the non-observable
monitoring condition, ceteris paribus) and SEEs derived via the delta
method from mixed-effects ordered logit and ordered probit regressions
with group-level random intercepts of a two-by-two treatment indicator
on the count of high effort choices. The $p$-value of a Wald test
of the null hypothesis that the respective estimate is equal to zero
is reported in parentheses. Rejections are ``highly significant''
for $p<.01$, ``significant'' for $p<.05$, and ``marginally significant''
for $p<.1$. Both regressions have $1,320$ observations clustered
in $88$ groups (i. e. $15$ rounds per group) and use mean-variance
adaptive Gauss--Hermite quadrature integration. Limit log-likelihood
was $-838.35149$ in the logit and $-837.32576$ in the probit model.
Likelihood-ratio tests reject the respective pooled specification
without random effects ($\bar{\chi}^{2}\left(01\right)=765.98$ and
$p=.0000$ for logit, $\bar{\chi}^{2}\left(01\right)=768.52$ and
$p=.0000$ for probit). The joint Wald tests reject the null model
($\chi^{2}\left(3\right)=8.09$ and $p=.0441$ for logit, marginally
for probit at $\chi^{2}\left(3\right)=7.58$ and $p=.0555$). Additional
session-level random intercepts are close to zero and not significant.}{\scriptsize\par}
\end{table}

\newpage
\section{Experimental Instructions \label{subsec:Instructions}}

\textbf{Screen 1: Welcome}\\

Welcome to the Birmingham Experimental Economics Laboratory. This is an experiment in decision making.\\

 The University of Birmingham has provided funds for this research. Just for showing up you have already earned \pounds{2.50}.\\
 
 You can earn additional money depending on the decisions you will make in today's experiment. It is therefore very important that you \textbf{read these instructions with care}.\\
 
 It is also important that you \textbf{remain silent} and do \textbf{not look} at other people's work.\\

If you have any questions, or need assistance of any kind, please raise your hand and an experimenter will come to you.\\

If you talk, laugh, exclaim out loud, etc., you will be asked to leave and you will not be paid. We expect and appreciate your following of these rules.\\

You will \textbf{first go over the instructions}, which are shown on-screen. After you have read the instructions, you will have time to \textbf{ask clarifying questions}.\\

We would like to stress that any choices you make in this experiment are \textbf{entirely anonymous}.\\
\newpage
\textbf{Screen 2: Instructions 1}\\

 Overview\\
 
 The experiment consists of \textbf{two parts}. In the \textbf{first part} you will make decisions in \textbf{15 identical tasks ("rounds")}. In the \textbf{second part} you will only make \textbf{one} single decision. In total you will thus make 16 decisions. At the end of the experiment one out of these sixteen decisions will be selected randomly (with equal probability) to determine your income from the experiment. Each decision is therefore equally important for determining your income. You will be paid \textbf{in private and in cash}.\\
 
 Part 1: The rules for rounds 1 through 15 will be identical and will be explained to you in detail below.\\
 
 Part 2: There will be additional rules for part 2 that will be explained to you before you make any decision in this part. The task in part 2 is similar, but not identical, to the task in part 1. The way you decide in the first part has no influence on the rules of the second part. You will receive new instructions for part 2 once everyone in the room has completed part 1.\\
 
 The next screens contain a detailed description of the rules.\\
 
 \newpage
 \textbf{Screen 3: Instructions 2}\\
 
 \textbf{Details}\\
 
 \textbf{Matching}\\
 
 At the beginning of the experiment, you will be anonymously matched with one other person, randomly selected from the participants in this room, to form \textbf{a group of two ("pair")}. Each person in the group will be assigned a role, either \textbf{ "Participant A"} or \textbf{"Participant B"}. It is equally likely for a person in a group to be either Participant A or Participant B.\\
 
 The group composition will remain the same during all 15 rounds of part 1. You will therefore interact \textbf{exactly 15 times} with the same participant in your group. For part 2 you will be matched with a new participant, but you will remain in the same role.\\
 
\textbf{ Decisions}\\

First, Participant A must \textbf{choose between two alternatives} represented by two different colors: \textbf{yellow} or \textbf{pink}. The payoff consequences of this choice are described on the next screen.\\

After Participant A made her or his decision, Participant B must make two consecutive decisions[\textbf{Exogenous} Participant B must make her or his decision]:\\

[\textbf{Endogenous}] First, Participant B must choose whether she or he \textbf{wants to find out} about the previous choice (between yellow and pink) of Participant A, prior to making her or his own decision.\\ 
 
[\textbf{Exogenous}]  First, \textbf{the computer} will determine if Participant B \textbf{will find out} about the previous choice (between yellow and pink) of Participant A, prior to making her or his own decision. The probability with which the choice of Participant A will be revealed is not known to either participant, but is based on average choices made by \textbf{previous participants} of this experiment, who have participated in a different version at an earlier date. \textbf{Different} from the version you are participating in, in this version Participants B \textbf{could decide} whether to look at the choices of Participant A prior to making their own choice. Note that at the end of a round, Participant A {\b will NOT be informed} about.\\
 
[\textbf{Intransparent Monitoring}] Note that at the end of a round, Participant A \textbf{will NOT be informed} about whether participant B has decided to find out about participant A's choice or not.[\textbf{Exogenous:} whether Participant B was informed by the computer about Participant A's choice or not]\\

[\textbf{Transparent Monitoring}]  Note that at the end of a round, Participant A \textbf{will be informed} about whether participant B has decided to find out about participant A's choice or not.[\textbf{Exogenous:} whether Participant B was informed by the computer about Participant A's choice or not]\\

Second, Participant B also must \textbf{choose between two alternatives}: \textbf{yellow} and \textbf{pink}.\\

Participant B makes her or his choice either knowing or not knowing the choice of the other participant, depending on her or his first decision.[\textbf{Exogenous} depending on what the computer has selected.]\

 \textbf{Screen 4: Instructions 3}\\

 \textbf{Details}\\
 
 \textbf{Earnings}\\
 
 The earnings for both participants from a given round are determined as follows:\\
 
  - If participant A and participant B  \textbf{both choose yellow}, they will earn \textbf{\pounds{8} each}.\\
  - If participant \textbf{A chooses yellow} and participant \textbf{B chooses pink}, participant\textbf{ A will earn \pounds{0}} and participant \textbf{B will earn \pounds{10}}\\
    - If participant \textbf{A chooses pink} and participant \textbf{B chooses yellow}, participant\textbf{ A will earn \pounds{10}} and participant \textbf{B will earn \pounds{0}}\\
      - If participant A and participant B  \textbf{both choose pink}, they will earn \textbf{\pounds{2} each}.\\

\textbf{End of round}

After each round, there is a \textbf{feedback screen}. On this screen \textbf{no further decisions} are made. Participant A will see the choice of participant B and the resulting payoffs. Similarly, Participant B will see the choice of participant A and the resulting payoffs.\\

[\textbf{Endogenous Transparent}] Additionally, Participant A will be informed of whether Participant B has decided to find out about Participant A's decision, before making her or his own choice between yellow and pink.\\

[\textbf{Exogenous Transparent}] Additionally, Participant A will be informed of whether Participant B was informed by the computer about Participant A's decision, before making her or his own choice between yellow and pink or was uninformed.\\

\textbf{ End of part 1}\\

All 15 rounds of part 1 will be played according to these rules.\\

You will receive additional instructions for part 2 after the first ten rounds. Your choices in part 1 will not affect the rules of part 2.\\

Do you have any questions? Please raise your hand and an experimenter will come to your desk. Please do not ask any question out loud.\\

 \textbf{Screen 5: Briefing}\\
 
 Your role is A(B)\\
 
 On this screen you are informed about your role (A or B) in the experiment that has been assigned randomly to you. You will now be matched with another participant that has the respective other role. Please press "OK" to continue\\

  \textbf{Screen 6: Decision A}\\
 
 Your role is A\\
 
 Your decision\\
 Radiobutton Pink\\
 Radiobutton Yellow\\
 
You must choose between two alternatives, either alternative "yellow" or alternative "pink". The consequences of your choice for your income depend on the choice of the other participant in your group, as explained in the instructions. You make your choice by clicking the respective box. You submit your choice by clicking the "OK" button. You can revise your choice as long as you did not yet submit it, after submission you cannot change your choice anymore.\\

\textbf{Only Endogenous Screen 7:  Monitoring Choice B}\\
 
 Your role is B\\
 
Do you want to be informed about the choice of participant A before you make your choice between "yellow" and "pink"?\\
 Radiobutton Yes\\
 Radiobutton No\\
 
[\textbf{Intransparent}] Participant A will \textbf{not be informed} about the choice you make here.\\
 
[\textbf{Transparent}] Participant A will \textbf{be informed} about the choice you make here.\\
 
On this screen you must choose between two alternatives, "no" and "yes". If you select "yes" you will be informed on the next screen, before you make your choice between "yellow" and "pink", about whether participant has chosen "yellow" or "pink". If you select "no" you will not be informed about the choice of participant A before you make your choice between "yellow" and "pink". You make your choice by clicking the respective box. You submit your choice by clicking the "OK" button. You can revise your choice as long as you did not yet submit it, after submission you cannot change your choice anymore.\\

\textbf{Screen 8: Decision B}\\
 
 Your role is B\\
 
 [\textbf{If Monitoring Chosen/Picked}: Participant A has chosen Yellow(Pink)]\\
 
 Your decision\\
 Radiobutton Pink\\
 Radiobutton Yellow\\
 
If you have chosen to be informed about the choice of participant A ("yellow" or "pink") on the previous screen (\textbf{Exogenous: If the computer has randomly chosen to inform you about the choice of participant A }), this information is shown in the mid-screen box. Otherwise, a question mark is shown in that box. You must then choose between two alternatives as well, either alternative "yellow" or alternative "pink". The consequences of your choice for your income depend on the choice of the other participant in your group, as explained in the instructions. You make your choice by clicking the respective box. You submit your choice by clicking the "OK" button. You can revise your choice as long as you did not yet submit it, after submission you cannot change your choice anymore.\\

\textbf{Screen 8: Feedback Screen A}\\

Your Choice: Yellow (Pink)\\

The choice of the other person in your group: Yellow (Pink)\\

Your payoff:\\

[\textbf{Transparent Endogenous}] Participant B has (not) gathered information on your choice prior to making his own choice.\\

[\textbf{Transparent Exogenous} The computer has (not) provided information on your choice to Participant B before he could make his own choice.\\

On this screen no further decisions are made, but the results of the current round are summarized. Once both participants in the group clicked the "OK" button the current round ends and the next round begins if the current round is not the last one.\\

\newpage

\textbf{Screen 9: Feedback Screen B}\\

Your Choice: Yellow (Pink)\\

The choice of the other person in your group: Yellow (Pink)\\

Your payoff:\\

On this screen no further decisions are made, but the results of the current round are summarized. Once both participants in the group clicked the "OK" button the current round ends and the next round begins if the current round is not the last one.\\

\textbf{Screen 10: Instructions Part 2}\\

Part 2 will be based on the same task and the same payoff rules as in part 1. It simply differs in the procedures by which participants can select their preferred option. Instead of playing consecutively, we are asking each participant \textbf{to state a plan of play} beforehand that specifies their choices conditional on the behaviour of the other participant in their group.\\

 In this part you will be \textbf{re-matched with a new participant} in this room. This means that you will not interact with the same participant as in part 1, but you will have the \textbf{same role} as before. However, it is important for your understanding of the rules that you \textbf{read the following instructions for both roles}.\\

This screen contains information on part 2 of the experiments. If you have any questions please raise your hand. An experimenter will come to your desk. Please do not ask your question aloud.\\

\newpage
\textbf{Screen 11: Instructions Part 2}\\

\textbf{ Participant B:} Instead of making your decisions successively as in part 1, you now must specify a complete decision plan for the entire task beforehand, conditional on the possible choices of participant A (and before knowing the actual choice of participant A). That is, you must choose \textbf{one from the following six plans:}\\

\begin{enumerate}
\item Don't find out about Participant $A$'s choice and choose yellow 
\item Don't find out about Participant $A$'s choice and choose pink 
\item Find out about Participant $A$'s choice and choose yellow independently of $A$'s choice 
\item Find out about Participant $A$'s choice and choose pink independently of $A$'s choice 
\item Find out about Participant $A$'s choice and choose pink if $A$ chooses pink, and yellow if $A$ chooses yellow 
\item Find out about Participant $A$'s choice and choose yellow if $A$ chooses pink, and pink if $A$ chooses yellow 
\end{enumerate}

After selecting your preferred plan, we will also ask you about your expectations of the choices of participant A. You can earn additional money if you correctly anticipate her or his choice. These procedures will be explained to you on a separate screen in more detail.\\

\textbf{Screen 12: Instructions Part 2}\\

\textbf{ Participant A:} You must also specify your choices conditional on the plan of participant B. That is, you  \textbf{choose between yellow and pink for any of the six plans} participant B could possibly choose (and before knowing the actual choice). Thus, you make \textbf{six decisions} in total. Only one of these decisions will determine your actual earnings, contingent on which of the six plans participant B selects.\\

After selecting your preferred choices, we will also ask you about your expectations of the choices of participant B. You can earn additional money if you correctly anticipate her or his choice. These procedures will be explained to you on a separate screen in more detail.\\
\end{document}